# Exploration of the Ice Giant Systems

A White Paper for NASA's Planetary Science and Astrobiology Decadal Survey
2023-2032

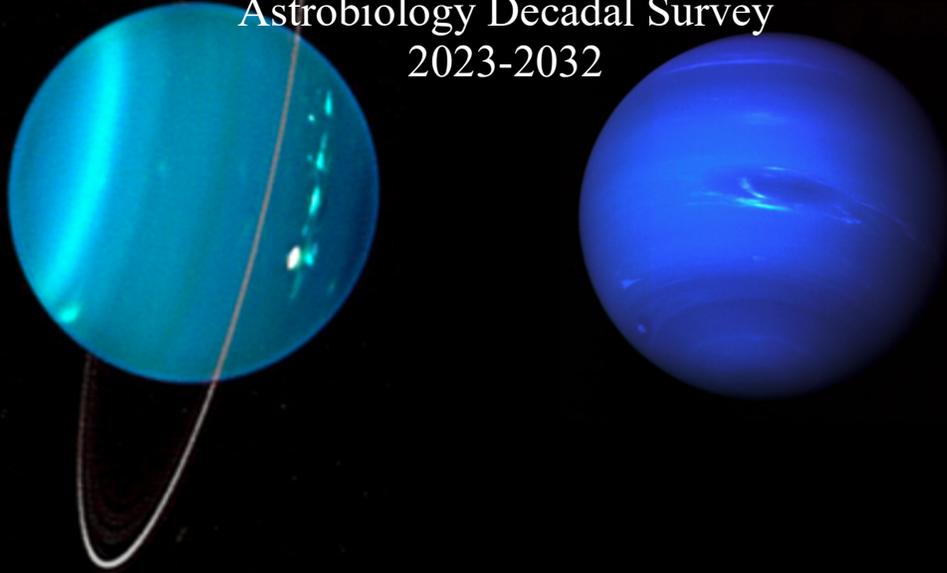

*Uranus (left) [1] and Neptune (right) (NASA)*

**Lead Authors:**

*Chloe B. Beddingfield*[1,2]
[1]The SETI Institute
[2]NASA Ames Research Center
chloe.b.beddingfield@nasa.gov

*Cheng Li*[3]
[3]University of California, Berkeley
cli@gps.caltech.edu

**Primary Co-Authors:**

| | | |
|---|---|---|
| Sushil Atreya[4] | Patricia Beauchamp[5] | Ian Cohen[6] |
| Jonathan Fortney[7] | Heidi Hammel[8] | Matthew Hedman[9] |
| Mark Hofstadter[5] | Abigail Rymer[6] | Paul Schenk[10] |
| Mark Showalter[1] | | |

[4]University of Michigan, Ann Arbor, [5]Jet Propulsion Laboratory, [6]Johns Hopkins University Applied Physics Laboratory, [7]University of California, Santa Cruz, [8]Association of Universities for Research in Astronomy, [9]University of Idaho, [10]Lunar and Planetary Institute

**Additional Coauthors and Endorsers:**
For a full list of the 136 additional coauthors and endorsers, see pages 8-10.

**Motivation**

> Ice giants are the only unexplored class of planet in our Solar System. Much that we currently know about these systems challenges our understanding of how planets, rings, satellites, and magnetospheres form and evolve. We assert that an ice giant Flagship mission with an atmospheric probe should be a priority for the decade 2023-2032.

Investigation of Uranus or Neptune would advance fundamental understanding of many key issues in Solar System formation: 1) how ice giants formed and migrated through the Solar System; 2) what processes control the current conditions of this class of planet, its rings, satellites, and magnetospheres; 3) how the rings and satellites formed and evolved, and how Triton was captured from the Kuiper Belt; 4) whether the large satellites of the ice giants are ocean worlds that may harbor life now or in the past; and 5) the range of possible characteristics for exoplanets.

Several time-critical science objectives require an ice giant mission in the coming decade. For example, a mission to Uranus would need to arrive no later than the early 2040s, because the unimaged northern high latitudes of the planet and five satellites will still be visible before the system transitions into southern spring in 2049. Additionally, the hypothesis that sunlight on Triton's south polar cap drives plume activity would become difficult to investigate around the 2050s, after the Neptune system has transitioned into southern fall (see white paper by [2]). Furthermore, optimal launch opportunities using a Jupiter gravity assist exist in the 2030-2034 timeframe for Uranus, and in the 2029-2030 timeframe for Neptune [3]. Both the Uranus and Neptune systems are compelling scientific targets, yet key differences exist between them. Both systems must ultimately be explored to achieve a complete understanding of the ice giant planet class, which is crucial for understanding how the Solar System formed and evolved. Exploration of the ice giants was a top recommendation in the previous two Decadal Surveys, and a mission to Uranus or Neptune with an atmospheric probe is endorsed by OPAG in a white paper [4].

**Origin, Evolution, and Interior Structures**

> Most planetary formation models have difficulty producing ice giants. Investigating the compositions and structures of ice giant interiors will provide information on how this type of planet, which appears to be common in the galaxy, formed and evolved.

*How do ice giants form?* The Voyager flybys of Uranus and Neptune more than three decades ago and ground-based observations have revealed that the ice giants, Uranus and Neptune, belong in a class of their own, distinct from the gas giants, Jupiter and Saturn. Conventional formation models − core accretion and gravitational instability − do not satisfactorily predict the properties of Uranus and Neptune. The so-called "runaway accretion" phase should have rapidly transformed them into gas giants [5]. Present models of the origin of the ice giant planets range from a slow process of formation in the neighborhood of Jupiter and Saturn followed by migration to their present orbital locations to rapid formation at their present orbits before the primordial solar nebula dissipated. However, observational constraints to test these extreme scenarios are unavailable. The evolution of the ice giants over the past 4.6 Ga is another unsolved mystery, which manifests itself in such quantities as the planetary heat balance, which is starkly different at Uranus compared to Neptune. Addressing the origin and evolution of the ice giants requires knowledge of the bulk atmospheric composition.

The abundances of heavy elements (mass greater than helium) determined from the bulk composition provide critical observational constraints to the formation and evolution models of the ice giant planets. In particular, the noble gases He, Ne, Ar, Kr, and Xe, and their isotopic ratios, are key, and only entry probes are capable of measuring them. Stable gas isotopic ratios, especially $^{13}C/^{12}C$, $^{34}S/^{32}S$, and D/H provide additional constraints. Knowledge of the bulk elemental abundances of C, S, N, and O are less important for understanding the formation of the ice giants than they are at Jupiter, but knowing the distribution of their bulk reservoirs ($CH_4$, $H_2S$, $NH_3$, and $H_2O$) to the deepest atmospheric level possible is valuable for understanding atmospheric dynamics and the interior structure, including a possible water ocean and an ionic/superionic water ocean at tens to hundreds of kilobar levels.

*What are the bulk compositions and interior structures of the ice giants?* While the planet's bulk densities are suggestive of water and other fluid ices being a dominant component, standard adiabatic 3-layer models typically have ice:rock ratios of 3:1 to 20:1 [6], which are all higher than expected from solar abundances, or from the composition of large icy moons or Pluto (which is 70% rock). These ratios strongly suggest that assumptions inherent in these standard models are incorrect, raising fundamental questions. For example, do high-pressure mixtures of high-density rock and low-density H/He mimic the density of ice? Are there discrete compositional layers in the interior, or are the rock, ice, and potentially gas more mixed? What is the heavy-element enrichment in the H/He envelope? (For more details, see the white paper by [7].)

*Are the ice giant interiors fully convective?* There are two significant reasons why there may be barriers to convection within these planets, which would drive the interior to higher temperatures than that expected from adiabatic calculations. One is the possibility of composition gradients at many locations, including the deep atmosphere due to cloud condensation at tens to hundreds of bars, and deeper interior gradients at the interfaces of H/He and ices, and ices and rocks. The second is the confirmed superionic phase of water, for which the oxygen acts as a solid, while the hydrogen acts as a fluid. The mixing allowed by this phase, and its effects on the geometry on the inferred dynamo region, are still unknown.

**Atmospheres**

The ice giant atmospheres are distinct from the gas giants in terms of their driving energy sources, their circulations, and their compositions. Understanding the ice giants will advance our knowledge of fundamental atmospheric processes and of planetary formation and evolution.

*What is the energy source that drives ice giant atmospheric activities?* Atmospheric motions can be forced by internal heat and solar insolation, which have a dramatic impact on circulation. If internal heat dominates, the eddy motion of the atmospheric currents would have enough kinetic energy to sufficiently homogenize tracers in the atmosphere like $CO_2$. However, if solar insolation dominates, heavy molecules will separate from the $H_2$/He envelope and produce a density stratification due to gravity. Redistributed heavy molecules and the absorbed heat adjust the density of the atmosphere and circulation. Therefore, determining the overall energy balance of the atmosphere (i.e., how much internal heat enters it from below) and identifying energy sources at different levels of the atmosphere, is crucial for understanding the fundamental processes governing atmospheric circulation and the distribution of tracer species. Uranus and Neptune represent extremes regarding internal heating and solar insolation and challenge our theories of atmospheric circulation [8]. Among all four giant planets, the internal/solar heating ratio is largest for Neptune (1.6), and smallest for Uranus (near zero).

*What is the pattern of the general circulation of ice giant atmospheres?* Determining the zonal, meridional, and vertical flow in the atmosphere, and the nature of storms and atmospheric waves, helps us understand how all giant planet atmospheres organize themselves and evolve. For example, we would like to know why Uranus and Neptune both have strong retrograde equatorial jets, while Jupiter and Saturn's are prograde, and why the poles of both ice giants appear to have strong mechanically forced downwellings [9] in spite of differences in their energy balances and the distribution of solar energy (Uranus' obliquity of 98° means its poles receive more sunlight on an annual average than the equator).

*What is the composition of the atmosphere?* As discussed previously, the composition of the atmosphere contains critical clues about the formation and evolution of the entire planet. The horizontal and vertical distribution of species, both condensable and not, also offers clues about the circulation and chemistry of the atmosphere. For example, are both ice giant atmospheres depleted by a factor of ~100 in $NH_3$ gas relative to solar abundances, as inferred from radio observations, while other species such as $CH_4$ and $H_2S$ are strongly enriched [10]? Our experience at Jupiter and Saturn, and planetary formation models, suggest all those "ice" species should have similar enrichments relative to solar. Is this a sign of chemical interactions with a deep ionic ocean?

**Rings and Small Satellites**

Investigating the ring systems and small moons of Uranus and Neptune will provide new key knowledge of the processes operating in astrophysical disks and the origin and history of solid material around the ice giants. The compositions of these objects around Neptune would provide insight into the material present before the capture of Triton.

*What processes sculpt the current system of ice giant rings and small moons into their current configurations?* Uranus hosts a system of dense narrow rings that lacks obviously meaningful spacing, diverse broad and finely-structured dusty rings, and the most tightly-packed system of small moons in the Solar System [11]. In contrast, Neptune has one ring composed of material organized into a series of longitudinally-confined arcs, its own suite of both broad and narrow dusty rings, and several moons that orbit interior to the most prominent rings [12]. These ring systems differ significantly from those of the gas giants, and thus can provide novel insights into the processes that can sculpt and maintain astrophysical disks under a variety of conditions.

*How does ring material evolve over time?* Both of the ice giant ring systems exhibit temporal variability. The locations and widths of Uranus' narrow rings oscillate in various ways on orbital timescales, while the location of one of Uranus' dusty rings appears to have changed over the course of several decades. The brightness of Neptune's ring arcs can also change dramatically over the years. Detailed measurements of these structures over an extended time would clarify the physical processes involved. Also, because some of Uranus' small moons are likely to collide with each other over short timescales [13] while Neptune's innermost moons are at risk of being torn apart by tidal forces [14], both of these systems could exhibit material cycling between rings and moons [15], yielding insights into aggregation and fragmentation processes.

*What are the compositions and origins of ice-giant rings and inner small moons?* Neptune's rings and small moons are probably remnants of the material present before Triton was captured [16]. Hence the composition of these bodies can provide information about the solid material that surrounded Neptune when it formed. However, the composition of Uranus' rings and small moons are different from its larger moons and from each other: generally, the rings' near-infrared spectra are flat whereas Uranus' larger moons have $H_2O$ ice and $CO_2$ ice spectral features [17], while certain moons are embedded in dusty rings with a variety of particle populations. The surface composition of the rings and small moons may therefore depend upon their present environment as well as their origins. Both the surface and the bulk compositions of most of the rings and small moons are not well constrained because spectral data are very limited [18] and there are few estimates of the moons' mass densities [19]. More detailed information about the compositions of the rings and small moons would provide new insights into the evolution of the Uranus and Neptune systems.

**Satellites**

> The large moons of Uranus and Neptune are possible ocean worlds with some surprisingly young surface regions. Investigation of these bodies would enhance our knowledge of where potentially habitable bodies exist in our Solar System.

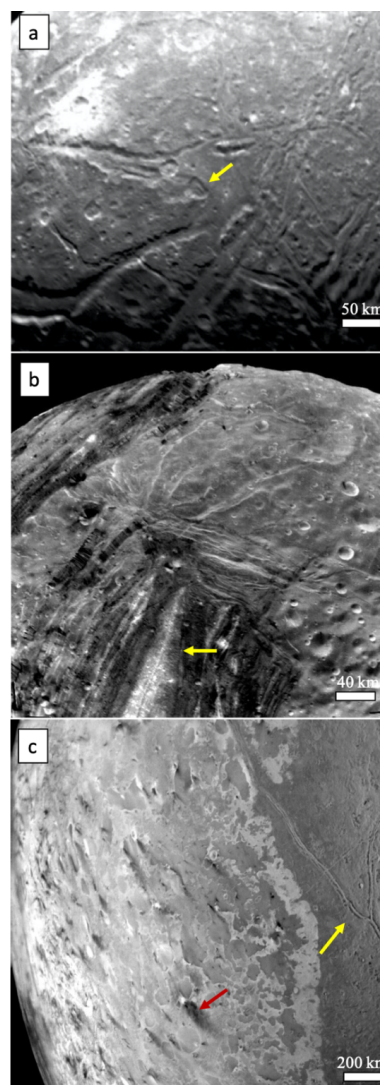

*Figure 1:* a) Ariel. A crater that may partially infilled by cryolava. b) Miranda. Tectonized terrain with albedo variations. c) Triton. Double ridges (yellow) and plume deposits (red).

***Which satellites are ocean worlds?*** The five classical Uranian satellites and Neptune's moon Triton may be ocean worlds [20] that display evidence for recent geologic resurfacing, including cryovolcanic activity [21] and high internal heat [22-24]. For example, the smooth floors of Ariel's chasmata exhibit lobate features that may have formed from flow of liquid water sourced from the interior, termed cryolava flows (Fig. 1a), and parallel high-standing ridges that are separated by topographic lows, reminiscent of fissure-style volcanism on Earth [20]. Miranda exhibits impact craters with subdued rims, reminiscent of the likely plume-mantled craters on the ocean world Enceladus [25], hinting at a possibly similar plume-driven mantling process [26], and complex fractured and faulted terrains are present (Fig. 1b). Umbriel exhibits evidence for resurfacing [27] and the large crater Wunda has a bright annulus of material that may have resulted from cryovolcanic infilling [28] (white paper by [32]).

Neptune's moon Triton is likely a captured Kuiper Belt dwarf planet with properties similar to Pluto, allowing for an important comparison. Triton exhibits double ridges with morphologies like those found on the ocean world Europa [29], and active plumes [30], indicating that the cryovolcanism is a prominent process on this world (Fig. 1c). Furthermore, the surface compositions of the ice giant moons are rich in volatile constituents ($CO_2$ ice, likely $NH_3$-bearing species, $N_2$, $CH_4$, and $CO$), suggesting recent activity and seasonal volatile migration [e.g., 31]. To determine if the large ice giant satellites are ocean worlds, investigations of induced magnetic fields, plumes and cryovolcanic features, surface heat anomalies, and inventories of volatile constituents are paramount (see white paper by [32]).

***What geologic processes modify the satellites and is there communication between the surfaces and interiors?*** Miranda, Ariel, and Titania show terrain with tectonic features, including fractures and faults, indicative of complex histories of geologic activity (Fig. 1a, b) [33, 34]. Triton is a captured Kuiper Belt Object, which would have caused it to experience substantial heating and complex tidal effects. Triton exhibits abundant evidence for recent and ongoing geologic activity [35], a tenuous atmosphere, and cantaloupe terrain that may have formed from upwelling of subsurface material, termed diapirism (see white paper by [2]). These widespread geologic features on the moons of Uranus and Neptune are indicative of recent and probably ongoing internal activity, and their investigation is crucial to gain insight into the processes operating on these worlds (see white papers by [2], [32], and [36]).

**Magnetospheres**

Investigating Uranus' and Neptune's uniquely oriented and variable magnetospheres is critical to advance our understanding of how magnetospheres are configured, populated, and evolve.

***How do magnetic field orientations relate to the configurations and dynamics of planetary space environments?*** Uranus and Neptune have uniquely asymmetric magnetic fields [37] (Fig. 2) and their dramatic orientations relative to the planets' rotational axes result in complex magnetospheric configurations that vary wildly on both diurnal and seasonal timescales [38]. These configurations are crucial missing data points to address how planetary properties relate to the configuration and dynamics of planetary space environments.

***Why does Uranus have electron radiation belts similar in intensity to the Earth's?*** The space environments surrounding the ice giants are also unique in that they are "vacuum magnetospheres" that are largely devoid of magnetospheric plasma. It is believed that such plasma is also the source for the radiation belts that encircle magnetized planets. It remains unknown whether these magnetospheric states observed by Voyager 2 are representative of these systems or a snapshot that captured reconfigurations.

***Do magnetospheres far from their host stars shed mass and couple to the solar wind?*** A magnetosphere needs to constantly shed the material produced within it. Yet such a process was not observed at Neptune by Voyager 2. At Uranus, such a process was observed, suggesting that the system is solar-wind-driven, which is curious and raises yet another question: if Uranus is solar-wind-driven, where is the solar wind material? In order to resolve these mysteries, we need the spatially- and temporally-distributed exploration of the plasma, radiation, fields, and wave distributions, as well as the aurora, of these planets that can only be provided by orbiting spacecraft, as discussed in a white paper [39].

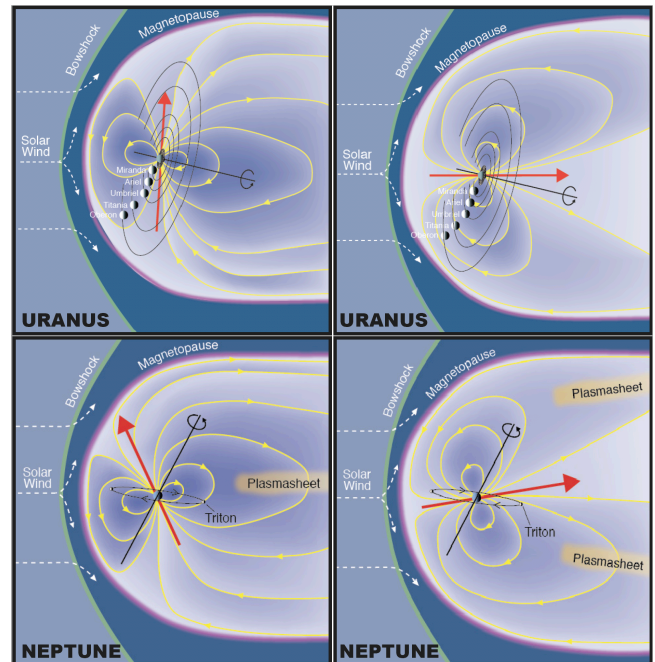

*Figure 2: The highly-tilted magnetic fields produce unique configurations with respect to the solar wind that vary drastically on diurnal and seasonal timescales. This may affect magnetospheric dynamics, satellite weathering, atmospheric coupling and auroral processes. Image credit: Bagenal and Bartlett.*

**Comparative Planetology among Solar System Gas Giants and Exoplanets**

Exoplanets with sizes comparable to our ice giants are common, and *in situ* observations of Uranus or Neptune would dramatically enhance our understanding of this planet class. Additionally, investigating the differences between ice- and gas-giant systems would provide novel insights into the processes that shape giant planet systems.

Our understanding of gas giant systems was revolutionized by orbiting spacecraft such as Galileo, Juno, and Cassini; orbiters would do the same for the ice giants. Furthermore, ice giant sized planets are abundant among the discovered extrasolar planets. Although most of these exoplanets are 'hot Neptunes' that orbit close to their host stars, ice giant planets likely represent a substantial fraction of existing exoplanets, and better understanding our own Solar System's ice giants is critical for understanding these exoplanets. With the scheduled launch of the James Webb Space Telescope (JWST), the infrared emission of numerous exoplanets will be available and a new era of characterization of their atmospheric conditions will ensue. For the most populous large exoplanets to be calibrated against ground truth from Solar System planets, it will be crucial to have *in situ* observations of Uranus or Neptune.

Because our Solar System does not host super-Earths or sub-Neptunes, Uranus and Neptune may be the best local representatives of the mid-sized planets that populate our local neighborhood. But, despite their importance, our knowledge of Neptune and Uranus atmospheres, compositions, and internal structures are poorly constrained. As discussed above, we know that local ice giant atmospheres significantly differ from those of the gas giants. For example, both Uranus and Neptune have retrograde jets near the equator while both gas giants have prograde jets. The ice giants show a pronounced difference in thermal emission between the poles and equator which are not present on gas giants. Unlike the gas giants, the ice giants are depleted in $NH_3$ gas. Water and hydrogen are well mixed in the interior of gas giants while this may not be the case for ice giants. Additionally, the magnetic fields of ice giants exhibit complex multiple polar structures, while those of the gas giants are mostly dipolar.

**Science at Uranus versus Neptune**

> Uranus and Neptune are equally compelling science targets, and ultimately, it will be necessary to explore both before we can claim to understand ice giants as a class of planet.

Uranus and Neptune have similar origins but notably different characteristics, and the study of each system would provide different insights into the nature of ice giants as well as giant planets in general, and the formation of our Solar System. For example, a mission to the Uranian system would allow us to observe the only giant planet in the Solar System with an atmosphere that is not strongly heated from the interior, and to explore a system that experienced an event significant enough to have resulted in its large tilt. The inner rings and small moons of Uranus allow us to explore one of the most dynamically chaotic regions in the Solar System, with significant changes occurring on decadal scales. The Uranus system provides the opportunity to study native Ice Giant moons that may be ocean worlds. In contrast, a mission to Neptune would allow us to investigate the captured KBO dwarf planet Triton, considered to be an ocean world with active plumes. Neptune's other icy worlds may reflect the Triton capture event.

**Technology Development**

> A radioisotope power system is the only new technology required for an ice giant mission.

Ice-giant missions can be accomplished with no new technology, except the development of a long-lived, next generation radioisotope power system [3]. Other technologies, some of which are well on the path to being developed or can be modified from commercially-available sources, can reduce costs and enhance the science return. Examples include the following: using advanced launch vehicles; aerocapture; cryogenic propellants; low mass/power/volume avionics and instruments designed for SmallSats. With the vast distance to the ice giants, large deployable Ka-band antennas (~5-meters) would increase the amount of data returned and autonomous systems can enable efficient scientific planning and scheduling and increase the amount and value of scientific data returned. All these would serve to maximize scientific return while reducing cost.

**Programmatic Considerations**

> Optimal launch opportunities for trajectories to the ice giants using a Jupiter gravity assist are in the upcoming decade, during the 2029-2034 range.

The above discussion demonstrates that the science to be done at the ice giants is compelling, which is the primary justification for a Flagship mission to an ice giant system. Programmatic considerations also favor such a mission. Regarding programmatic balance, ice giants are the only class of planet never to have had a dedicated mission. Furthermore, an ice giant Flagship is itself well-balanced and broad in its goals: it will engage all disciplines within the planetary science community, targeting the interior, atmosphere, rings, satellites, potential ocean worlds, and the magnetosphere; it also engages the exoplanet community. Few missions can advance such a broad range of disciplines (even if the Trident Discovery proposal is selected, its target is Triton, not the Neptune system). Considerations of cost sharing and international cooperation also favor the ice giants, as ESA has expressed a strong interest in participating in a NASA-led ice giant mission.

Programmatic considerations can also help decide where to send the first ice giant mission. To fully understand this class of planet, we must visit both, but cost considerations may drive us to first target only one, in which case schedule, risk and (again) cost may weigh in. Optimal trajectories to Neptune occur in the 2029-2030 timeframe, while optimal Uranus trajectories are 2030-2034 [3]. Missions to either planet can be launched outside those windows, but with a penalty in either payload mass or risk. A Uranus mission is expected to cost $300M less than a comparable Neptune mission if launched in the optimal windows (ibid) although Neptune may become much more expensive if launched after 2031 because of the need for advanced launch vehicles such as NASA's SLS as well as the need for using radioisotope power supplies and spacecraft components beyond their nominal lifetimes (based on NASA's Planetary Mission Concept Studies program preliminary results). If launched within the optimal windows, missions to either planet are possible with current technologies, making them extremely high pay-off, low-risk [3], scientific endeavors.

# Additional Coauthors and Endorsers

| | | |
|---|---|---|
| 1. | Caitlin Ahrens | University of Arkansas |
| 2. | Karen Aplin | University of Bristol |
| 3. | Giada Arney | NASA Goddard Space Flight Center |
| 4. | Shahid Aslam | NASA Goddard Space Flight Center |
| 5. | Abigail Azari | University of Michigan |
| 6. | Kevin Baillié | Observatoire de Paris, France |
| 7. | Erika Barth | Southwest Research Institute, Boulder |
| 8. | Christopher Bennett | University of Central Florida |
| 9. | Tanguy Bertrand | NASA Ames Research Center |
| 10. | Ross Beyer | The SETI Institute / NASA Ames Research Center |
| 11. | Ali Bramson | Purdue University |
| 12. | Carver J. Bierson | Arizona State University |
| 13. | Marina Brozovic | Jet Propulsion Laboratory |
| 14. | Shawn Brueshaber | Western Michigan University |
| 15. | Emma Bruce | University of Leicester, UK |
| 16. | Devon Burr | Northern Arizona University |
| 17. | Bonnie Burratti | Jet Propulsion Laboratory |
| 18. | Paul Byrne | North Carolina State University |
| 19. | Xin Cao | University of Iowa |
| 20. | Richard Cartwright | The SETI Institute |
| 21. | Julie Castillo-Rogez | Jet Propulsion Laboratory |
| 22. | Rob Chancia | Rochester Institute of Technology |
| 23. | Nancy Chanover | New Mexico State University |
| 24. | George Clark | Johns Hopkins Applied Physics Laboratory |
| 25. | Andrew Coates | University College London, UK |
| 26. | Corey Cochrane | Jet Propulsion Laboratory |
| 27. | Geoff Collins | Wheaton College |
| 28. | Joshua Colwell | University of Central Florida |
| 29. | Frank Crary | University of Colorado |
| 30. | Dale Cruikshank | NASA Ames Research Center |
| 31. | Athena Coustenis | Paris Observatory at Meudon |
| 32. | Matija Ćuk | The SETI Institute |
| 33. | Emma Dahl | New Mexico State University |
| 34. | Ingrid Daubar | Jet Propulsion Laboratory |
| 35. | David DeColibus | New Mexico State University |
| 36. | Imke DePater | University of California, Berkeley |
| 37. | Rajani Dhingra | University of Idaho |
| 38. | Chuanfei Dong | Princeton University |
| 39. | Catherine Elder | Jet Propulsion Laboratory |
| 40. | Joshua Emery | Northern Arizona University |
| 41. | Anton Ermakov | University of California, Berkeley |
| 42. | Sierra Ferguson | Arizona State University |
| 43. | Leigh Fletcher | University of Leicester, UK |
| 44. | Robert French | The SETI Institute |
| 45. | Keenan Golder | University of Tennessee |
| 46. | Felipe González | University of California, Berkeley |
| 47. | Mitchell Gordon | The SETI Institute |
| 48. | Cesare Grava | Southwest Research Institute, San Antonio |
| 49. | Denis Grodent | Université de Liège, Belgium |
| 50. | Will Grundy | Lowell Observatory |
| 51. | Noah Hammond | Wheaton College |
| 52. | Emily Hawkins | Loyola Marymount University |
| 53. | Alex Hayes | Cornell University |
| 54. | Paul Helfenstein | Cornell University |
| 55. | Ravit Helled | University of Zurich |
| 56. | Douglas Hemingway | Carnegie Institution for Science |
| 57. | Amanda Hendrix | Planetary Science Institute |
| 58. | Chuck Higgins | Middle Tennessee State University |
| 59. | Amy Hofmann | Jet Propulsion Laboratory |
| 60. | Bryan Holler | Space Telescope Science Institute |
| 61. | Timothy Holt | University of Southern Queensland |
| 62. | George Hospodarsky | University of Iowa |
| 63. | Carly Howett | Southwest Research Institute, Boulder |
| 64. | Hsiang-Wen Hsu | University of Colorado |
| 65. | Caitriona M. Jackman | Dublin Institute for Advanced Studies |
| 66. | Jamie Jasinski | Jet Propulsion Laboratory |
| 67. | Devanshu Jha | MVJ College of Engineering |
| 68. | Xianzhe Jia | University of Michigan |
| 69. | Insoo Jun | Jet Propulsion Laboratory |
| 70. | Erich Karkoschka | Lunar and Planetary Laboratory |
| 71. | Simon Kattenhorn | University of Alaska, Fairbanks |
| 72. | Mallory Kinczyk | North Carolina State University |
| 73. | Michelle Kirchoff | Southwest Research Institute |
| 74. | Peter Kollmann | Johns Hopkins Applied Physics Laboratory |
| 75. | Kartik Kumar | Delft University of Technology |
| 76. | James Keane | California Institute of Technology |
| 77. | Sascha Kampf | University of Colorado |
| 78. | Erin Leonard | Jet Propulsion Laboratory |
| 79. | Cecilia Leung | Jet Propulsion Laboratory |
| 80. | Jack Lissauer | NASA Ames Research Center |
| 81. | Rosaly Lopes | Jet Propulsion Laboratory |
| 82. | Michael Lucas | University of Tennessee |
| 83. | Alice Lucchetti | Astronomical Observatory of Padova, Italy |
| 84. | Kathleen Mandt | Johns Hopkins Applied Physics Laboratory |
| 85. | Essam Marouf | San Jose State University |
| 86. | Emily Martin | Smithsonian Institute |
| 87. | Adam Masters | Imperial College London |
| 88. | Melissa McGrath | The SETI Institute |
| 89. | Tim I. Michaels | The SETI Institute |
| 90. | Karl Mitchell | Jet Propulsion Laboratory |
| 91. | Sarah E. Moran | Johns Hopkins University |
| 92. | Julianne Moses | Space Science Institute |
| 93. | Quentin Nenon | University of California at Berkeley |
| 94. | Marc Neveu | NASA Goddard Space Flight Center |
| 95. | Francis Nimmo | University of California Santa Cruz |
| 96. | Tom Nordheim | Jet Propulsion Laboratory |
| 97. | Maurizio Pajola | Astronomical Observatory of Padova, Italy |
| 98. | Robert Pappalardo | Jet Propulsion Laboratory |
| 99. | Chris Paranicas | Johns Hopkins Applied Physics Laboratory |
| 100. | Alex Patthoff | Planetary Science Institute |
| 101. | Carol Paty | University of Oregon |
| 102. | Michael Person | MIT Planetary Astronomy Laboratory |
| 103. | Georgia Peterson | University of British Columbia |
| 104. | Noemi Pinilla-Alonso | Florida Space Institute |
| 105. | Simon Porter | Southwest Research Institute, Boulder |
| 106. | Frank Postberg | Free University of Berlin |
| 107. | Alena Probst | Jet Propulsion Laboratory |
| 108. | Lynnae Quick | NASA Goddard Space Flight Center |
| 109. | Leonardo Regoli | Johns Hopkins Applied Physics Laboratory |
| 110. | Edgard G. Rivera-Valentín | Lunar and Planetary Institute (USRA) |
| 111. | Thomas Roatsch | German Aerospace Center |
| 112. | Aki Roberge | NASA Goddard Space Flight Center |
| 113. | James Roberts | Johns Hopkins Applied Physics Laboratory |
| 114. | Stuart Robbins | Southwest Research Institute, Boulder |
| 115. | Sébastien Rodriguez | Université de Paris |
| 116. | Joe Roser | The SETI Institute |
| 117. | Elias Roussos | Max Planck Institute, Germany |
| 118. | Kirby Runyon | Johns Hopkins Applied Physics Laboratory |
| 119. | Francesca Scipioni | The SETI Institute |
| 120. | Jennifer Scully | Jet Propulsion Laboratory |
| 121. | Kelsi Singer | Southwest Research Institute |
| 122. | Krista Soderlund | University of Texas |
| 123. | Mike Sori | Purdue University |
| 124. | Linda Spilker | Jet Propulsion Laboratory |
| 125. | Alan Stern | Southwest Research Institute, Boulder |
| 126. | Ted Stryk | Roane State Community College |
| 127. | Marshall J. Styczinski | University of Washington |
| 128. | Ali Sulaiman | University of Iowa |
| 129. | Daniel Tamayo | Princeton University |
| 130. | Jim Thieman | NASA Goddard Space Flight Center |
| 131. | Matthew Tiscareno | The SETI Institute |
| 132. | Paolo Tortora | University of Bologna, Italy |
| 133. | Elizabeth Turtle | Johns Hopkins Applied Physics Laboratory |
| 134. | Orkan Umurhan | The SETI Institute / NASA Ames Research Center |
| 135. | Benjamin Weiss | Massachusetts Institute of Technology |
| 136. | Oliver White | The SETI Institute / NASA Ames Research Center |